\documentclass{article}
\usepackage{spconf,amsmath,graphicx}

\usepackage{enumitem}
\usepackage{hyperref}
\usepackage[ruled,vlined]{algorithm2e}
\usepackage{graphicx}
\usepackage{amssymb,xcolor,hyperref}
\usepackage[numbers]{natbib}
\usepackage{subfig}
\usepackage[ruled,vlined]{algorithm2e}
\usepackage{amsmath}
\usepackage{soul}
\usepackage{algcompatible}
\usepackage{multirow}
\usepackage{setspace,lipsum}
\setlist{nosep, leftmargin=14pt}

\usepackage{mwe} 

\title{An Embarrassingly Simple Consistency Regularization Method for Semi-Supervised Medical Image Segmentation}
\name{Hritam Basak$^{1*}$, Rajarshi Bhattacharya$^{1\dagger}$, Rukhshanda Hussain$^{1\dagger}$, Agniv Chatterjee$^{1\dagger}$ 
\address{$^1$ Dept. of Electrical Engg., Jadavpur University, Kolkata, India-700032}
\thanks{$^\dagger$ Equal contribution; $^*$ Corresponding Author}}

\address{}
\begin{document}
%
\maketitle
\begin{abstract}
The scarcity of pixel-level annotation is a prevalent problem in medical image segmentation tasks. In this paper, we introduce a novel regularization strategy involving interpolation-based mixing for semi-supervised medical image segmentation. The proposed method is a new consistency regularization strategy that encourages segmentation of interpolation of two unlabelled data to be consistent with the interpolation of segmentation maps of those data. This method represents a specific type of data-adaptive regularization paradigm which aids to minimize the overfitting of labelled data under high confidence values. The proposed method is advantageous over adversarial and generative models as it requires no additional computation. Upon evaluation on two publicly available MRI datasets: ACDC and MMWHS, experimental results demonstrate the superiority of the proposed method in comparison to existing semi-supervised models. Code is available at: \hyperlink{https://github.com/hritam-98/ICT-MedSeg}{https://github.com/hritam-98/ICT-MedSeg}      
\end{abstract}


\section{Introduction}
\label{intro}
Supervised medical image segmentation is a widely explored problem in computer vision, achieving exponential growth recently. These methods mostly rely upon deep learning, requiring large-scale pixel-wise annotation data. However, acquiring huge amounts of labelled medical data is tedious and expensive, thus methods alleviating this requirement are highly expedient. Semi-Supervised Learning (SSL) is a promising direction in this regard, requiring only a few labelled data and compensating for the large portion of unlabelled data by generating pseudo labels. Recently, SSL-based methods have been widely recognized for their superior performance in medical image segmentation. They not only eliminate the necessity of large-scale annotations but also produce accurate segmentation results that are very close to those obtained from supervised models.

Recently, Bai \emph{et al.} \cite{bai2017semi} proposed an iterative SSL framework where pseudo labels are generated after every iteration and refined using conditional random field (CRF). Adversarial learning is another popular method for utilizing unlabelled data. Zhang \emph{et al.} \cite{zhang2017deep} proposes a deep adversarial network (DAN) that encourages the segmentation of unlabelled data to be similar to those of labelled data. The state-of-the-art SSL segmentation model mean-teacher network is recently extended by Yu \emph{et al.} \cite{yu2019uncertainty} for uncertainty-aware left atrium segmentation.  

Consistency regularization is another plausible solution in this direction, that encourages realistic perturbations of an unlabelled image to produce consistent segmentation maps. For example, Bortsova \emph{et al.} \cite{bortsova2019semi} proposes a transformation-consistent segmentation network, capable of exploring the equivariance of elastic perturbations for precise lung X-ray segmentation and Li \emph{et al.} \cite{li2020transformation} proposes a semi-supervised segmentation framework using transformation consistency, encouraging consistent predictions of the network-in-training for different perturbation of the same input. Unlike these methods, which rely upon the low-density region assumption, our proposed method chooses perturbation directed towards another unlabelled sample, thereby reducing the necessity of expensive gradient calculation.  

On the other hand, interpolation-based regularizers have been achieving state-of-the-art performance in various tasks and across multiple architectures. Recently, this idea has been extended to an unsupervised setting by Berthelot \emph{et al.} \cite{berthelot2018understanding} where the authors propose that utilizing realism of the latent space interpolation from autoencoder can improve model learning. Driven by this speculation and the aforementioned success of consistency regularization in SSL methods, in this paper we investigate an interpolation-based mixing technique in a semi-supervised setting and its utility in medical image segmentation. 
\section{Methodology}\label{sec:method}
\begin{figure}
    \centering
    \includegraphics[width=\columnwidth]{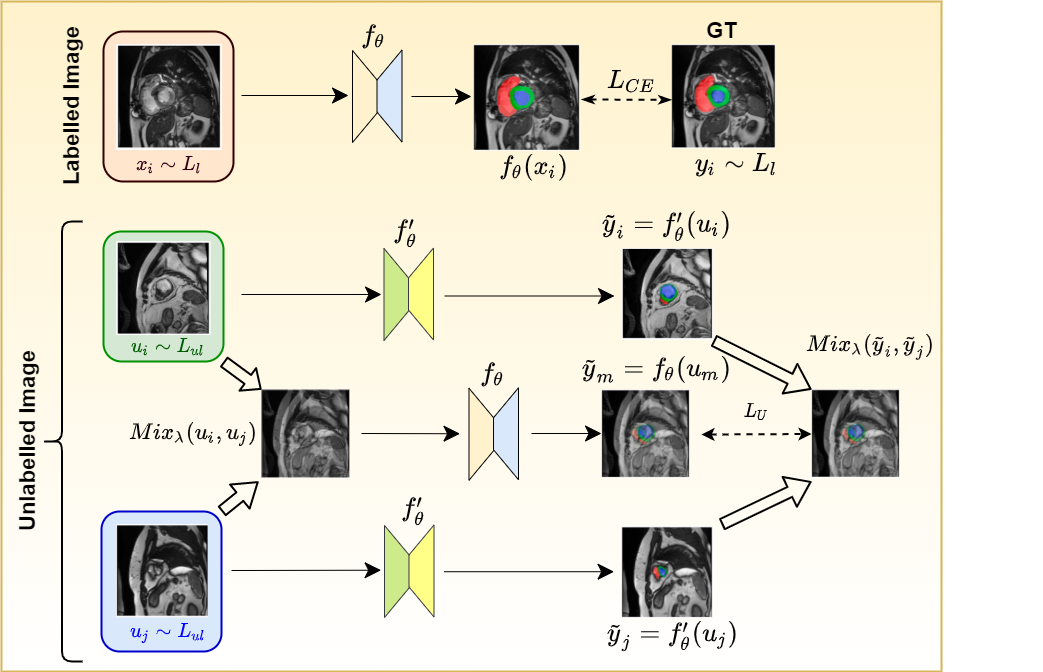}
    \caption{Overall framework of our proposed architecture.}
    \label{overall_diag}
\end{figure}

\subsection{Consistency Regularization with Interpolation}
Consistency Regularization has been used in existing literature as a means to enhance the robustness of a network pipeline by enforcing the network against various perturbations on the unlabelled data, to increase the generalizability of these new data points. Among them, the most effective perturbation would be in the adversarial direction - in the direction almost perpendicular to the decision boundary between the positive and negative examples, i.e., the direction in which the network is most liable to misclassify the pixels. However, most existing literature incorporates perturbations that may or may not be in the adversarial direction, and thus, results in loss of generalizability. Some other methods do perturb the input in the adversarial direction but require a large amount of unlabelled data and thus, might not be feasible in the biomedical domain.

Thus, we have proposed a pixel-wise data perturbation strategy as consistency regularization in our work, which operates as described. The network is trained in such a way so as to ensure stable and accurate segmentation of image points interpolated from existing points. Considering two unlabelled image data-points ${u_1}$ and  ${u_2}$, we interpolate another unlabelled image point $\mathcal{M}_\alpha (u_1, u_2)$, where ${\mathcal{M}_\alpha (u_1,u_2) = \alpha u_1 + (1-\alpha) u_2}$, for some hyperparameter $\alpha$. Now, consistency regularization is applied between the output of the interpolated image data-point $f(\mathcal{M}_\alpha (u_1,u_2)$ and the interpolation of the outputs of the original unlabelled points $\mathcal{M}_\alpha (f(u_1),f(u_2)) = \alpha f(u_1) + (1-\alpha) f(u_2)$. 
This exploits the fact that the network learns to predict a pixel-level segmentation mask of the input images, and further, consistency is maintained between the outputs of the interpolated inputs and the interpolated outputs of the original inputs.

Thus, the unlabelled samples in the datasets are used to generate the new interpolated images and the corresponding pseudo-labels. It takes two unlabelled images as input and returns the interpolated image and the corresponding pseudo-label, which is used by the network pipeline. Therefore, the Consistency Regularization technique can be summarized as:
\begin{equation}
    \mathcal{M}_\alpha(f_{\theta '}(u_1), f_{\theta '}(u_2)) \simeq f_\theta(\mathcal{M}_\alpha (u_1, u_2))
\end{equation}


This data mixing technique would help the model learn more robust features improving the semi-supervised learning on subsequent (target) tasks since random perturbations do not guarantee adversarial perturbation.

\subsection{Objective Function}
Consider labelled samples $(x_i,y_i) \sim L_l$ from joint distribution $P(X,Y)$ and unlabelled samples $(u_i,u_j) \sim L_{ul}$ from borderline distribution $P(X)=\frac {P(X,Y)}{P(X|Y)}$. Using SGD for every iteration $t$, the encoder-decoder parameter $\theta$ is updated minimising the objective function:
\begin{equation}\label{loss1}
    \mathcal{L} = \mathcal{L_{CE}} + r(t).\mathcal{L_U}
\end{equation}
where $\mathcal{L_{CE}}$ is the cross entropy loss applied over the labelled data $L_l$ and $\mathcal{L_U}$ is the interpolation consistency regularization loss applied over the unlabelled data $L_{ul}$, $r(t)$ is the ramp function adjusting the weight of $\mathcal{L_U}$ after every iteration. $\mathcal{L_U}$ is calculated over $(u_i,u_j)$ of sampled minibatches and the pseudo labels $\tilde{y}_i=\mathcal{F}_{\theta'}(u_i)$ and $\tilde{y}_j=\mathcal{F}_{\theta'}(u_j)$ ($\theta'$ is the exponential moving average of $\theta$). Next, interpolation $u_m = {\mathcal{M}}_\alpha(u_i,u_j)$ and model prediction $\tilde{y}_m=\mathcal{F}_{\theta}(u_m)$ are computed updating $\theta$ to bring $\tilde{y}_m$ closer to the interpolation of the pseudo labels, $\mathcal{M}_\alpha(\tilde{y}_i,\tilde{y}_j)$. The deviation in $\tilde{y}_m$ and ${\mathcal{M}}_\alpha(\tilde{y}_i,\tilde{y}_j)$ is penalised using the mean squared loss. 
Therefore, $\mathcal{L_U}$ can be expressed as: 
\begin{equation}
\begin{aligned}
\label{loss2}
         \mathcal{L_U} = E_{u_i,u_j \sim L_{ul}}  l(\mathcal{F}_\theta({\mathcal{M}}_\alpha(u_i,u_j)),\\
    {\mathcal{M}}_\alpha(\mathcal{F}_{\theta'}(u_i),\mathcal{F}_{\theta'}(u_j)))
\end{aligned}
\end{equation}
Our overall approach is depicted in \autoref{overall_diag} and \autoref{algo}.
\begin{algorithm}
    {\small
    {\em Input:}\\
    $L_l$: Distribution of labelled samples; $L_{ul}$: Distribution of unlabelled samples\\ 
    {\em Define:}\\
    $f_\theta(\cdot)$: Segmentation network with trainable parameter $\theta$\\
    $f_{\theta '(\cdot)}$: Segmentation network with parameter $\theta '$- exponential moving average of $\theta$\\
    $T$: Total number of iterations; $r(t)$: ramp function \\
    $\lambda$: exponential moving average change rate\\
    $\mathcal{M}_\alpha(u_1, u_2) = \alpha u_1+(1-\alpha)u_2$
    \begin{algorithmic}
        \STATE\WHILE{(t $\leq$ T)}
            \STATE Sample labelled mini-batch $\rhd\: \{(x_p, y_p)\}_{p=1}^P \sim L_l$
            \STATE Supervised CE loss $\rhd\: \mathcal{L}_{\mathcal{CE}}(\{(f_\theta(x_p), y_p)\}_{p=1}^P)$
            \STATE Sample two unlabelled batches $\rhd\: \{(u_i, u_j)\}_{k=1}^U \sim L_{ul}$
            \STATE Generate pseudo-labels $\rhd\: \{\tilde{y_i}, \tilde{y_j}\}_{k=1}^U=\{f_{\theta '}(u_i, u_j)\}_{k=1}^U$
            \STATE Interpolation $\rhd\: u_m=\mathcal{M}_\alpha(u_i,u_j), y_m=\mathcal{M}_\alpha(\tilde{y_i}, \tilde{y_j})$
            \STATE Interpolated pseudo-label $\rhd\:\tilde{y}_m=f_\theta(u_m)$
            \STATE Unsupervised loss $\rhd\:\mathcal{L}_{\mathcal{U}} = MSE(\{y_m, \tilde{y}_m\}_{m=1}^U)$
            \STATE Overall model loss $\rhd\: \mathcal{L}=\mathcal{L}_{\mathcal{CE}}+r(t)\mathcal{L}_{\mathcal{U}}$
            \STATE Gradient computation $\rhd\: \mathcal{G}_\theta \longleftarrow \mathcal{L}\cdot\nabla _\theta$
            \STATE Update parameter $\rhd\: \theta ' \longleftarrow (1-\lambda)\theta+\lambda \theta '$
            \STATE $\theta \longleftarrow step(\mathcal{G}_\theta, \theta)$
         \ENDWHILE
         
    \end{algorithmic}
    return $\theta$\\
       
    }
\caption{Pseudo-code of our proposed method.}
\label{algo}
\end{algorithm}


\section{Experiments and Results}
\label{results}
\begin{table*}[htbp]
\centering
\caption{Comparison of the proposed method with State-of-the-art methods on the ACDC 2017 Dataset and MMWHS dataset.}
\label{tab:results}
\resizebox{0.8\textwidth}{!}{
\begin{tabular}{|c|c|c|c|c|c|c|}
\hline
\multirow{2}{*}{Methods} & \multicolumn{3}{c|}{ACDC 2017} & \multicolumn{3}{c|}{MMWHS} \\ \cline{2-7} 
 &
  \begin{tabular}[c]{@{}c@{}}labelled data\\ (as \% of the training set)\end{tabular} &
  DSC(\%) &
  HD (mm) &
  \begin{tabular}[c]{@{}c@{}}labelled data\\ (as \% of the training set)\end{tabular} &
  DSC(\%) &
  HD(mm) \\ \hline
Peng et al.\cite{peng2021boosting}              & 5\%     & 85.76     & -     & 13.30\%   & 55.75   & -     \\ \hline
Mix-Up\cite{zhang2017mixup}                  & 10\%     & 86.3      & -        & 50\%      & 79.6    & -     \\ \hline
Chaitanya et al\cite{chaitanya2020contrastive}          & 10\%     & 88.6      & -        & 50\%      & 79.4    & -     \\ \hline
Adversarial Training\cite{zhang2017deep}     & 20\%     & 79.1      & 5.16       & 50\%      & 77.9    & 3.20    \\ \hline
\textbf{Ours }                    & \textbf{10\%}     & \textbf{89.8}      & \textbf{4.47}     & \textbf{40\% }     & \textbf{79.83}   & \textbf{3.05}  \\ \hline
\end{tabular}
}
\end{table*}

\subsection{Dataset and Experimental Setup}

The experiments for evaluating the model performance was carried out on two public datasets: the ACDC 2017 \cite{bernard2018deep} dataset consisting of 100 short-axis cardiac MRI volumes with expert annotations for three structures: left and right ventricle and myocardium and the MMWHS \cite{zhuang2016multi} dataset, hosted in MICCAI 2017, consisting of 20 cardiac MRI samples with expert annotations for seven structures: left and right ventricle, left and right atrium, pulmonary artery, myocardium, and ascending aorta. The datasets were split into train, validation, and test sets, where 2 volumes were considered for validation purposes in both the datasets. The test set consists of 20 and 5 volumes for ACDC 2017 and MMWHS, respectively. The rest of the data was used for training. To study the segmentation performance with respect to the labelled data, we have experimented with 1.25\%, 2.5\%, and 10\% labelled volumes for ACDC 2017 and 10\%, 20\%, and 40\% labelled volumes for the MMWHS dataset, following the existing works in literature \cite{chaitanya2020contrastive}. We have used ResNet-50 as the encoder backbone of our architecture, using an ADAM optimizer with an initial learning rate of 1e-5. For evaluation purposes, we have used three widely used metrics: Dice Similarity Score (DSC), Average Symmetric Distance (ASD), and Hausdorff Distance (HD). Average of all the metric scores over all the classes and reported in this paper. Mixing parameter $\alpha$ was set experimentally.

\subsection{Results on the ACDC 2017 Dataset}

We achieve DSC of 73.56\%, 79.05\%, and 89.80\% on the ACDC 2017 dataset for 1.25\%, 2.5\%, and 10\% labelled volumes respectively. For the same set of experiments, ASD of 1.981, 1.546, and 1.229  and HD of 6.719, 5.820, and 4.473 were obtained respectively. The merits of the model can be better appreciated by referring to the comparison with other state-of-the-art methods, shown in \autoref{tab:results}, where we have compared our obtained results using 10\% labelled volumes only.
A higher Dice score at a lower percentage of labelled data indicates that the model is capable of generalizing the data efficiently by making use of a lesser number of labelled samples. As seen in the table, our approach uses a low number of labelled samples and produces the best results in terms of DSC and HD, whereas methods like \cite{zhang2017deep} use double the amount of labelled data, yet achieve inferior performance as compared to our results, proving the efficiency and robustness of our model. Only Peng \emph{et al.} \cite{peng2021boosting} uses a lower number of labelled data, but produces poorer results as well. Our method produces a reasonably better result by making use of only 10\% of the labelled data available for training.

\begin{figure}[tbp]
\begin{minipage}[b]{.48\linewidth}
  \centering
  \centerline{\includegraphics[width=4.0cm]{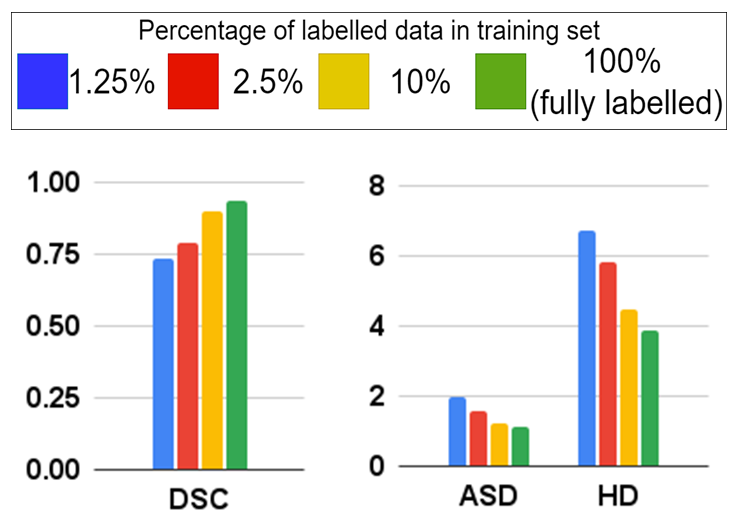}}
\end{minipage}
\hfill
\begin{minipage}[b]{0.48\linewidth}
  \centering
  \centerline{\includegraphics[width=4.0cm]{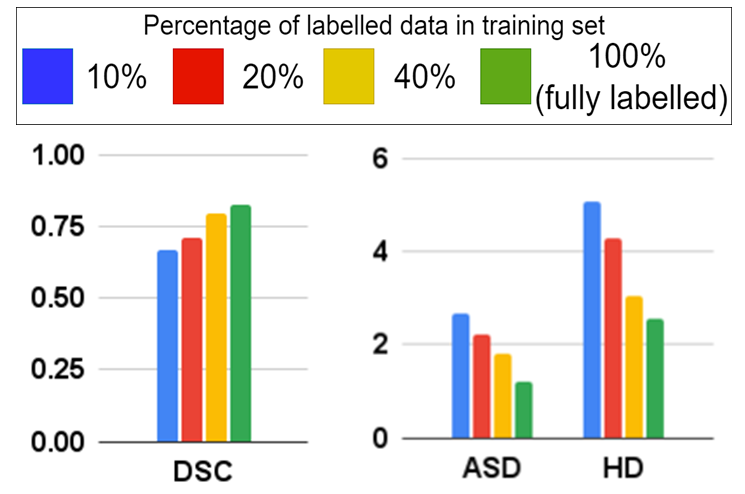}}
\end{minipage}
\caption{Comparison of results obtained by using different percentages of labelled data in training set on ACDC 2017 (left) and MMWHS (right) dataset.}
\label{comp_label}
\end{figure}

\subsection{Results on the MMWHS Dataset}
Upon evaluation on the MMWHS dataset, DSC of 66.86\%, 71.24\%, and 79.83\% have been achieved using 10\%, 20\%, and 40\% labelled volumes, respectively. ASD scores of 2.671, 2.207, and 1.819 and HD values of 5.073, 4.288, and 3.047 were achieved for these experiments, respectively. The efficacy of the proposed technique can be verified by comparison with existing state-of-the-art techniques as shown in \autoref{tab:results}, where we have compared our results obtained using 40\% annotated data only, yet achieving superior performance as compared to methods with $50\%$ labelled data. The results were reported from the papers directly. All the other parameters were the same as those methods, to maintain fair comparison.
\begin{figure}[tbp]
    \centering
    \includegraphics[width=0.85\columnwidth]{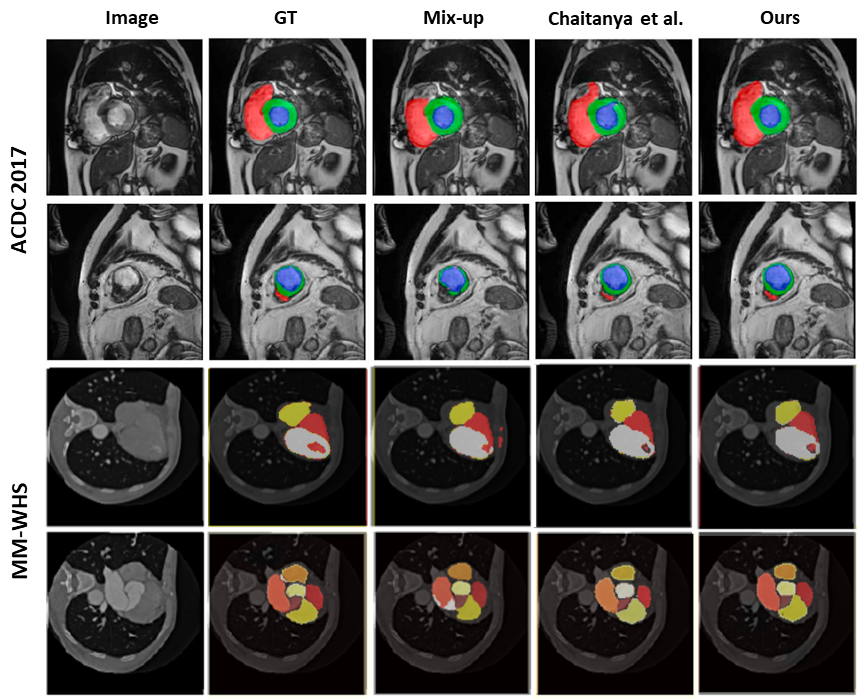}
    \caption{Comparative image of the segmentation masks of our model, along with the other semi-supervised models}
    \label{comparison_img}
\end{figure}
Peng \emph{et al.} \cite{peng2021boosting} use the least number of labelled samples in their method, but they have produced vastly substandard results. Whereas, our method obtains better results as compared to most state-of-the-art methods, by using 40\% labelled data in the entire training set. The results achieved by Zhang \emph{et al.} \cite{zhang2017mixup, chaitanya2020contrastive} are very close to ours, however, they have used a larger percentage of labelled data for training. 



\autoref{comp_label} represents the results obtained using different percentages of labelled data from the training set. It is obvious that employing more labelled data while training will produce superior performance. However, we can observe from the figure that our model can produce results very close to the supervised method (that uses 100\% labelled train data) by employing only a small percentage (10\% for ACDC 2017 and 40\% for MMWHS) of annotations. \autoref{comparison_img} offers a visual comparison of the proposed method with a few of the state-of-the-art methods existing in contemporary literature. A careful inspection of the image shows that our method generates labels that are the closest to the ground truth (GT) when compared to the other methods. ASD scores were not reported in most of the methods in the literature for both the datasets and hence not considered for comparison.

\section{Conclusion and Future Work}
\label{disc_conc}
Recently, machine learning has had a transformative impact on various applications, but its effectiveness is impacted due to the lack of labelled data, especially in medical imaging, where availing of those annotations is extremely difficult. Here we have tried to address this shortcoming by developing a semi-supervised learning strategy that encourages consistency regularization by interpolation. It is advantageous over the previous SSL models in multiple aspects: unlike adversarial perturbations or generative models, it requires almost no additional computations. Moreover, our proposed model outperforms several state-of-the-art SSL methods by utilizing fewer annotations than these methods. In future, we plan to extend the work by interpolating at the hidden representations to boost the performance and evaluate the performance in general computer vision problems to assess its robustness. We further plan to extend the experimentation by adding more variability between the two interpolating images.   

\section{Compliance with Ethical Standards}
This research study was conducted retrospectively using human subject data made available in open access. Ethical approval was \textbf{not} required as confirmed by the license attached with the open-access data.

\begin{spacing}{0.87}
\bibliographystyle{IEEEbib}
\bibliography{refs}
\end{spacing}

\end{document}